# Classification of Melanocytic Nevus Images using BigTransfer (BiT)

## A study on a novel transfer learning-based method to classify Melanocytic Nevus Images

Sanya Sinha (*Author*)
Dept. of Electronics and Communications Engineering
Birla Institute of Technology Mesra, Patna
Patna, India
ssanya0904@gmail.com

Nilay Gupta (*Author*)
Dept. of Computer Science and Engineering
Sikkim Manipal Institute of Technology
Majitar, India
ng.nilaygupta04@gmail.com

***Abstract*—Skin cancer is a fatal disease that takes a heavy toll over human lives annually. The colored skin images show a significant degree of resemblance between different skin lesions such as melanoma and nevus, making identification and diagnosis more challenging. Melanocytic nevi may mature to cause fatal melanoma. Therefore, the current management protocol involves the removal of those nevi that appear intimidating. However, this necessitates resilient classification paradigms for classifying benign and malignant melanocytic nevi. Early diagnosis necessitates a dependable automated system for melanocytic nevi classification to render diagnosis efficient, timely, and successful. An automated classification algorithm is proposed in the given research. A neural network previously-trained on a separate problem statement is leveraged in this technique for classifying melanocytic nevus images. The suggested method uses BigTransfer (BiT), a ResNet-based transfer learning approach for classifying melanocytic nevi as malignant or benign. The results obtained are compared to that of current techniques, and the new method's classification rate is proven to outperform that of existing methods.**

### I. INTRODUCTION

Skin cancer is a fatal disease that affects a massive population pool globally [1]. It can manifest itself in the form of lesions appearing on the epidermis [2] Early diagnosis of these lesions might facilitate speedy recovery [2]. For human skin color variation, Melanin is the important pigmentation factor. Because of the considerable resemblance between various forms of skin lesions, manual inspection is difficult and may result in incorrect investigation [3]. A dermatologist's accuracy percentage used to be between 65% and 80% based just on ocular assessment [4]. Dermatologists seldom attain sensitivities of higher than 80% [5]. Dermoscopy images were acquired with a powerful imaging equipment and camera lens in cases with suspected lesions along with the ocular assessment. During the recording process, inadequate lighting is adjusted using the application of filters to remove the anomalies in imaging the underlying dermal layers. With this added filter, the accuracy was boosted by 49% [6]. The overall accuracy of a dermatological examination is 75-84% [7-8]. As a result, for skin lesion classification, an automated method is necessary to improve both the speed as well as the accuracy of the diagnosis [9].

Former computer-enabled approaches for dermatological image classification had some fundamental flaws. The first of these was the shortage of adequate data [10-11]. The imaging procedure is the second issue, with skin images taken using specialized instruments and dermoscopy, and other medical images obtained using a microscope and biopsy [12]. To analyze the skin image data, previous approaches [13-14] needed substantial preprocessing, segmentation, and feature extraction operations.

Transfer learning (TL) is a machine learning methodology that focuses on preserving information obtained while solving one issue and applying it to another similar problem. In classical machine learning, we have different tasks and for each task, we have a different model. In transfer learning, we train a model for one kind of object and then use it for different kinds of objects of the same type. For example, we could train one model to recognise roses (a type of flower) and then retrain that same model to detect lotuses (another kind of flower).

In this paper, a novel TL method known as Big Transfer (BiT) is used to classify the colored images of melanocytic nevi into two classes: benign and malignant. BiT outperforms other methods in the automated diagnosis landscape for skin cancer over 2 parameters. First, this method is capable of working with both dermoscopic and photographic images. Second, very minor pre-processing and augmentation is required, unlike its counterparts that heavily depended on preprocessing. BiT only requires pre-training once, followed by inexpensive fine-tuning for down streaming. Other cutting-edge approaches, on the other hand, need significant training on data samples adapted to suit the given use-case [15-17]. BiT does not necessitate either a succinct fine-tuning approach or a hyperparameter tweaking methodology for every new task.

### II. RELATED WORK

Numerous cutting-edge techniques have been developed in the field of automated diagnosis due to the urgent requirement for the early diagnosis of melanoma. Among these, supervised learning-based techniques including Support Vector Machines (SVM) [18], K-means Clustering [19], and AdaBoost MC [20] for classifying dermoscopy images with multi-component patterns have been the most noteworthy. The need for the classification of benign and malignant lesions paved the way for techniques such as the ABCD rule with pattern recognition [21] and artificial neural networks [22]. Deep convolutional neural networks (DCNNs) for autonomous feature extraction were introduced by Esteva et al. [23], redefining the categorization of melanocytic lesions. CNNs significantly accelerated the diagnosis process and improved the classification of melanocytic cancers. A CNN-based approach for the binary classification of malignant and benign melanocytic nevi was proposed by Brinker et al. [24]. The outcomes were compared to those attained manually on 100 test images by 157 dermatologists. The CNN demonstrated the effectiveness of the model by outperforming 131 out of 157 dermatologists. Many related studies by Tschandl et al. [25] and Codella et al. [26] demonstrate how effective deep learning is at classifying medical images. As a data-hungry ensemble, CNNs need a lot of sources of data to train. Transfer Learning (TL) paradigms were created to address this issue. TL is a supervised learning method which imports the weights of a previously-trained model into a new model with a comparable problem statement. This improves computational efficiency and the model's performance in the more recent use case. In the past, a multitude of TL frameworks has been used to classify skin cancer-related images. ResNet50 [27], Inception V3 [28] and VGG16 [29] are just a few of the many DCNN frameworks employed to classify images diagnosed with multiple kinds of skin cancer.

We propose a TL method in this paper that leverages BigTransfer (BiT) to classify melanocytic nevi as benign or malignant for effective and speedy prognosis.

## III. Proposed Work

BigTransfer is a unique method for image classification proposed by Kolesnikov et al. [30] that leverages a transfer learning-based approach. The network is trained on millions of images from a generic dataset and the respective weights are then adjusted and fine-tuned to suit the target problem statement. BiT is trained on multiple, open-access datasets from the ImageNet challenge [31-34]. BiT achieves remarkably good results on each of these aforementioned datasets, thus paving the way for implementation on newer datasets.

BiT is a time-efficient network that needs to be pre-trained only once on a large dataset and then subsequently fine-tuned by adjusting the weights accordingly. It requires a short time period for fine-tuning the weights for every new problem statement and does not require any major tweaks in the hyperparameters. The methodology behind BiT has 2 phases in the training lifecycle: upstream pre-training, and downstream transfer.

### A. *Upstream Pre Training*

Scaling, Weight Standardization (WS) and Group Normalization (GN) are important pillars of upstreaming data from a voluminous, generic dataset to a dataset with relatively fewer data points. The robustness of the pre-training dataset is directly proportional to the complexity of the DCNN architecture leveraged for the task. Scaling is used to render the model adaptable to alien weights from the newer tasks. Scaling factors including model size, dataset size and training time are of critical importance. WS is a normalization technique revolving around micro-batch training. It is used to standardize the weights in the convolutional layers and smoothening the loss landscape by mitigating the values of the Lipschitz constants of loss and gradients. GN is a normalization layer that splits individual channels into groups and normalizes the features from each group. GN coupled with WS can be leveraged for macro-size batch training by outperforming complementary methods such as Batch Normalization (BN) significantly. BiT produces both unsatisfactory and computationally inefficient results with BN.

Before training, image preprocessing techniques such as random cropping and random horizontal mirroring are used, followed by a $224 \times 224$ image resize. The BiT models leverage a vanilla ResNet-v2 architecture wherein all BN layers are replaced by WS and GN layers. ResNet-152 architectures are trained across all generic datasets, with each hidden layer having a four-fold increase in width (ResNet152x4). All the models upstream are trained using standard gradient descent for optimization. The value of the momentum is kept at 0.9, while that of the initial learning rate (lr) is 0.03. The model training is performed for 50 epochs. Following this, a decay by a factor of 10 in the lr is done at the 20,30, and 40 epochs accordingly. The lr is subsequently multiplied by a factor of batch size/512. No weight decay is used during the transfer.

### B. *Downstream Fine-tuning*

BiT-HyperRule is a heuristic, fine-tuning methodology, created to filter and choose only the most critically important hyperparameters as an elementary function of the target image resolution and number of data points for model tuning. Training schedule length, resolution, and the likelihood of selecting MixUp regularization are a few notable selected hyperparameters. MixUp regularization delivered the best results in medium-sized datasets with BiT, and not for smaller-sized datasets. Most hyperparameters remain constant irrespective of the datasets, but parameters including resolution, schedule and the application of MixUp are expressed as functions of the image resolution and size of the training dataset sample. The SGD optimizer, loss function sparse categorical cross entropy and accuracy metrics were used while compiling the model.

### C. *Evaluation Metrics*

Top-1 Accuracy is the main evaluation metric used to evaluate the performance of BiT. For classification tasks, accuracy tells us what fraction of predictions were accurately classified by our model.

Accuracy is the main evaluation metric used to assess the model performance of BiT. For classification tasks, accuracy tells us what fraction of predictions were accurately classified by our model. The metrics observed from the data sample are TP (True Positive), TN (True Negative), FP (False Positive) and FN (False Negative).

$$Accuracy\ (\%) = \frac{TP + TN}{TP + TN + FP + FN} \times 100 \qquad (1)$$

Loss is a metric that is used to predict by what factor the returned value deviates from the target value. BiT works best with the sparse categorical cross-entropy loss function. It can be denoted mathematically as

$$J(w) = -\frac{1}{N}\sum_{i=1}^{N}[y_i \log(\hat{y}_i) + (1 - y_i)\log(1 - \hat{y}_i)] \qquad (2)$$

where,
w refers to the weights,
$y_1$ refers to the true label,
$\hat{y}_i$ is the predicted label

## IV. Results And Discussions

The BiT model was trained in a Python environment using the Keras API with the TensorFlow backend on an Intel(R) Core i7-10510U CPU with an NVIDIA Tesla P100 GPU containing 16GB of HBM2 vRAM.

### A. *Comprehensive Analysis on Model Validation*

The performance of the model is validated on 300 malignant and 360 benign test images. The performance of the model on the validation dataset is juxtaposed with that of cutting-edge deep learning techniques such as Inception V3 [28], VGG-16 [29], ResNet-50 [27], EfficientNetB0 [35] and machine learning techniques including SVM [18] and AdaBoost MC [20] in Table 1. The top accuracy for our model on the given dataset was a stunning 88.13%. Inception V3 had the next best accuracy of 84.43%, followed by VGG16 with 84.23% accuracy. While it is conventional for DCNNs to be better for image classification tasks, traditional methods like AdaBoost and SVM returned accuracies of 79.92% and 77.65% respectively. They outperformed popular DCNN frameworks like ResNet50 with 76.70% validation accuracy and EfficientNetB0 with 54.77% validation accuracy. It is observed from Fig.1. that the validation accuracy of the proposed methodology outperformed the respective performances of all of these state-of-the art methods. BiT does not overfit and returns a comparable accuracy of 89.2% on the training dataset as well. This illustrates minimalistic bias present in the training phase of the data.

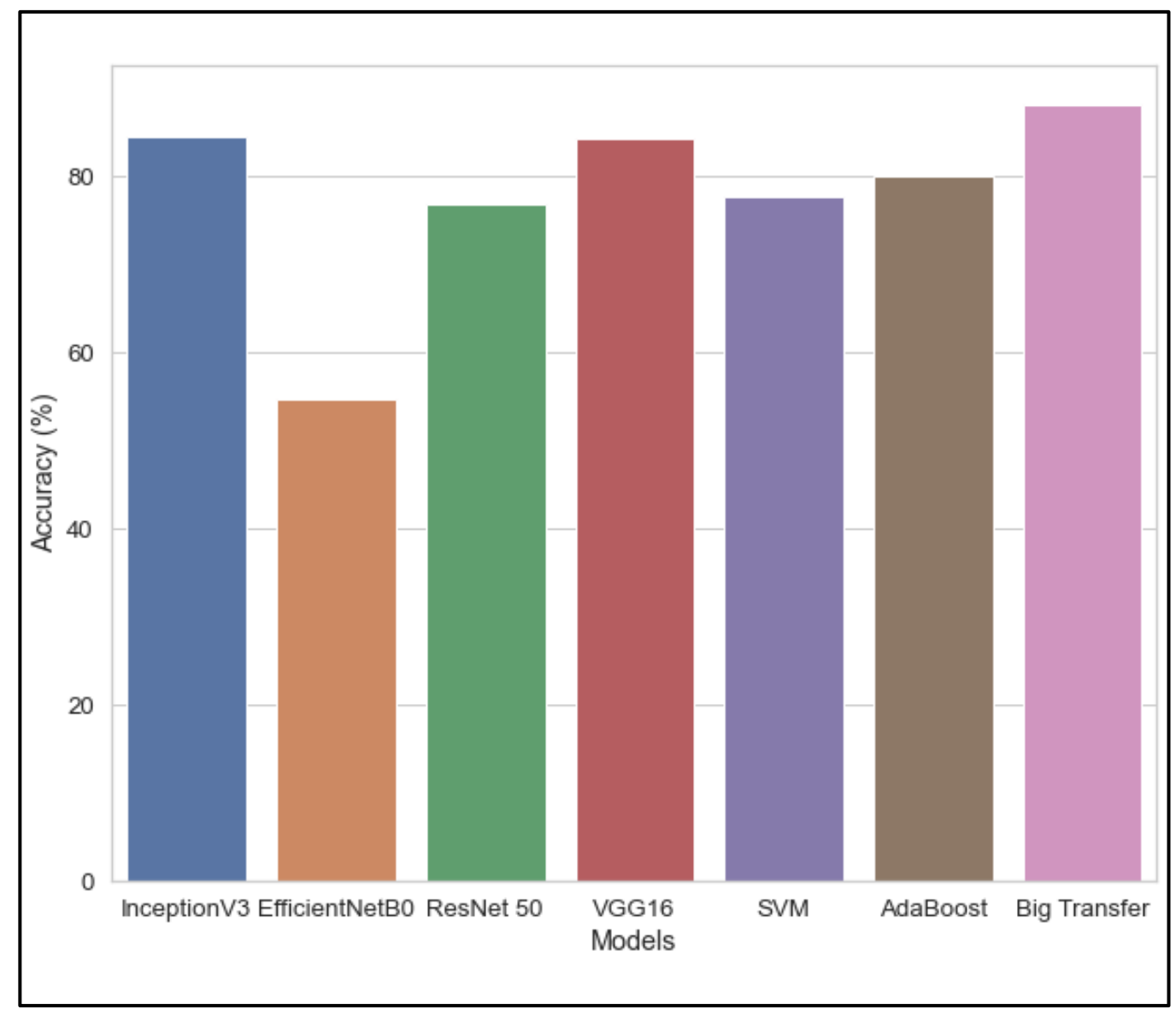

Fig. 1. Accuracy of all TL and ML models

### B. *Time Consumed for Model Training*

While BiT provides the best classification results, it is also the most computationally efficient. Fig.2. illustrates that it takes merely 40.32 seconds for training the BiT model. Also, the next best model, Inception V3, takes 555.38 seconds for the task, which is more than 13 times the time taken by BiT for a superior prediction. Machine learning methods, like SVM and boosting methods, like AdaBoost are the most time-consuming in this context. Other computationally efficient CNN-based techniques like EfficientNetB0 and ResNet50 do not provide satisfactory classification results. Thus, BiT can be the most accurate and computationally efficient method.

### C. *Accuracy and Loss Curves*

The performance of the model's accuracy and loss is evident in Fig.3. The training and validation accuracy and loss curves are plotted against the total number of epochs used while training. It can be observed that the training and validation accuracy are nearly equivalent towards the end of the training cycle. Similarly, the model's training and validation losses decline from nearly 0.4 to less than 0.2 towards the end of the training cycle. This increment in accuracies and decline in losses indicates a progressive model with optimized values of accuracies and losses. It is noted that there is not a broad gap between the training and validation sample set values of the accuracy and loss metrics. Hence, it can be mustered that our model does not overfit and returns quite resilient values.

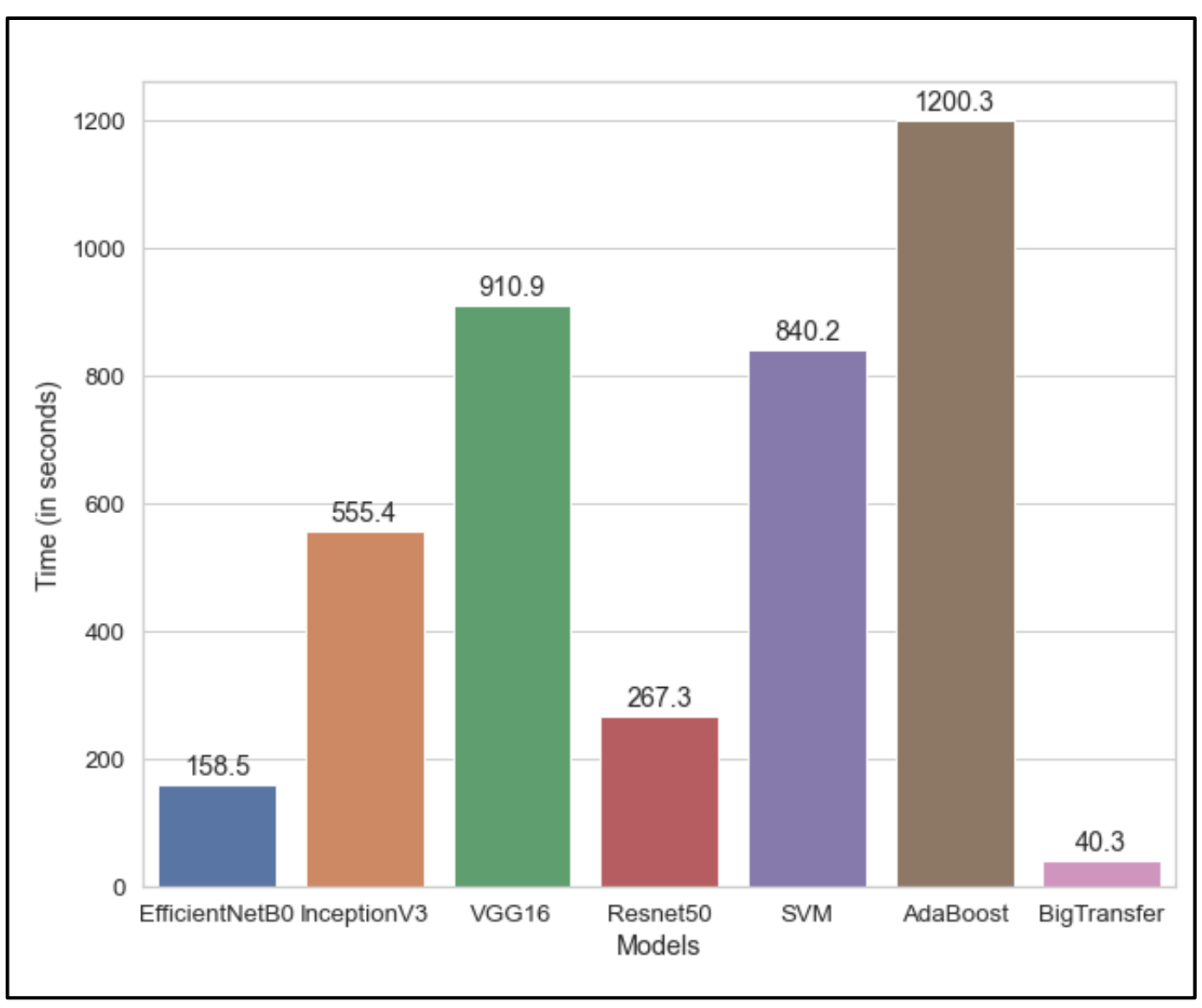

Fig. 2. Time taken to train (in seconds) of all TL and ML models

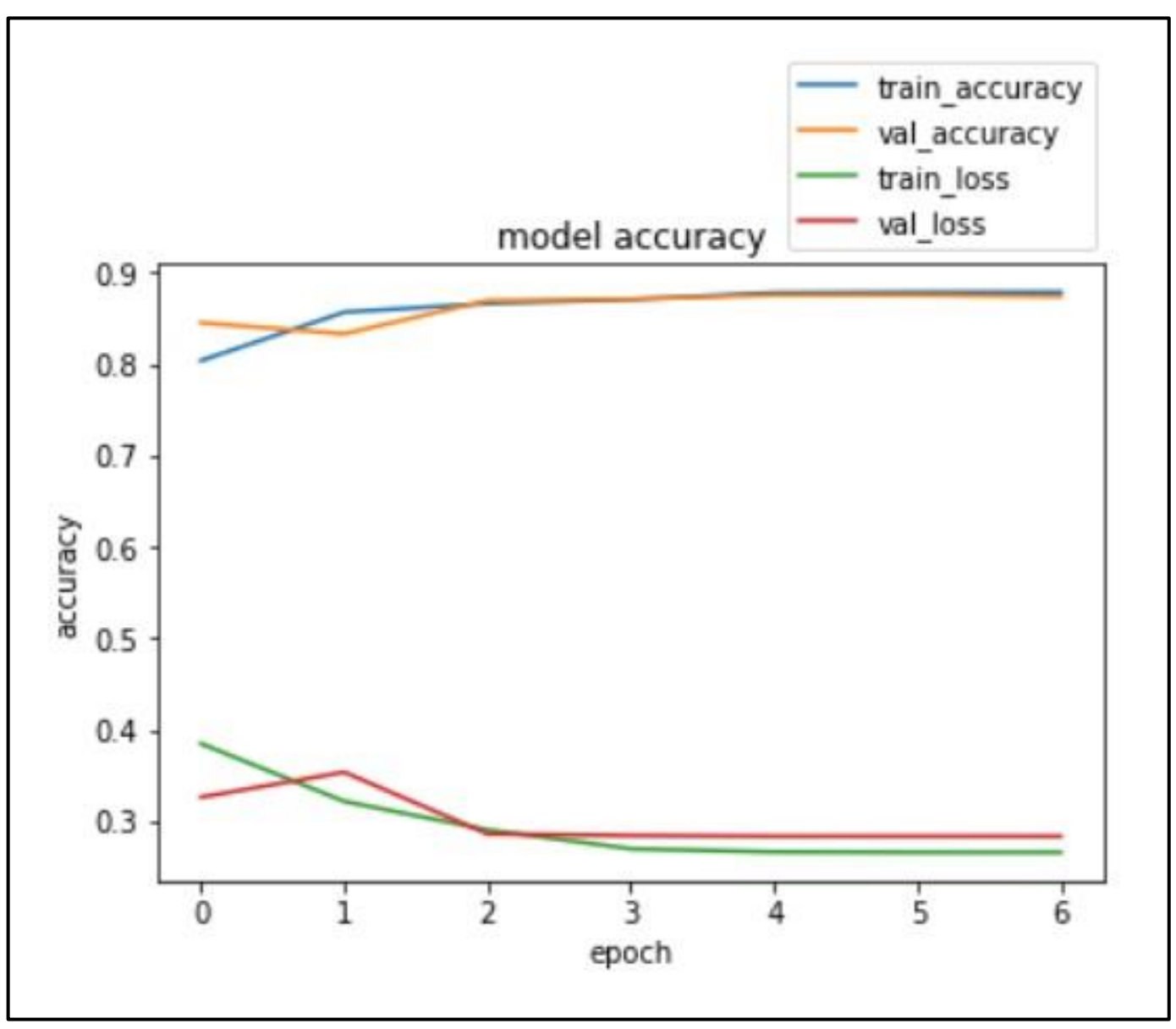

Fig. 3. Accuracy vs Loss for the Big Transfer (BiT) model

TABLE I. COMPARISON WITH OTHER WORKS

| Model Name | Inception V3 [28] | VGG 16 [29] | ResNet50 [27] | Efficient NetB0 [35] | SVM [18] | AdaBoost MC [20] | BiT (Proposed) |
|---|---|---|---|---|---|---|---|
| Accuracy (%) | 84.43 | 84.24 | 76.70 | 54.77 | 77.65 | 79.92 | **88.13** |

## V. CONCLUSION

In this paper, we have used BigTransfer (BiT) to classify melanocytic nevus images into benign and malignant. BiT is an extremely fast and resilient transfer learning-based method that is known for image classification tasks. In this paper, we have shown how BigTransfer can beat both state-of-the-art DCNN models like Inception V3, ResNet50 and VGG16, along with conventional machine learning models such as AdaBoost MC and SVM. BiT is particularly useful for visual representation learning. It is pre-trained on multiple, generic datasets, and is fine-tuned and adjusted to adapt to newer weights from a novel use-case. BiT is an upcoming, cutting-edge transfer learning approach that can outperform multiple, popular DCNN frameworks. It returns test results in 2-3 seconds maximum with commendable accuracy.

To popularize this potent neural network, we seek to use BigTransfer for multi-class classification on the HAM10000 [36] skin lesion dataset in the future. We also seek to reduce over-fitting in the training and validation data samples and leverage BigTransfer to the optimum extent for efficient and timely melanoma diagnosis.